# Finding merit in dividing neighbors

Kamran Behnia

LPEM(CNRS-UPMC)
ESPCI, 10 Rue Vauquelin
75005 Paris
FRANCE

A large fraction of energy consumed by humankind is wasted as heat. Some of this energy can be recycled with thermoelectric devices that convert a thermal gradient into electricity. However, their wide adoption will require development of materials with high thermoelectric figure of merit (ZT) that lack rare or harmful elements. In this issue, Zhao et al. (1) report on p-doped tin selenide (SnSe) helps meet these goals.

The thermoelectric figure of merit of a solid (ZT) is a dimensionless number quantifying its capacity to be used as an element in a thermoelectric generator or refrigerator. ZT is pushed up by enhancing Seebeck coefficient and electric conductivity and lowering thermal conductivity of the solid in question. For decades, researchers have looked for materials, which combine three distinct properties: large thermoelectric response, reasonable ability to conduct electricity and reluctance to let the heat travel. These are contradictory requirements and only a handful of materials have been found to meet them. Sadly, two of the two most prominent are $Bi_2Te_3$ (the champion near room temperature) and PbTe (the star close to 800 K), which contain rare tellurium or poisonous lead.

Last year, the same group identified SnSe as an interesting thermoelectric material with a large ZT near 900K thanks to its remarkably low thermal conductivity [2]. The new work reports on successful p-doping of the SnSe single crystals [1]. This accomplishment is accompanied with an increase in the magnitude of ZT, which, along one of the crystal axes, exceeds unity between 400 K and 800 K. Thus, SnSe competes with both $Bi_2T_3$ and PbTe over a wide temperature window.

The newcomer to this arena is a member of a family of binary salts, containing elements of column IV and column VI of the periodical table. The IV-VI family and column V elements crystalize in one of the three varieties derived from the cubic rocksalt structure (Fig.1) [3]. Tin selenide, like black phosphorous, prefers the orthorhombic crystal structure, the source of its anisotropic conductivity.

The complexity emerging in presence of only one or two type of atoms present has intrigued condensed-matter physicists for decades (3-5). Why should atoms opt for anything less symmetric than cubic? The rocksalt structure can be seen as a network of interpenetrating octahedra with an atom at the center of each octahedron and its six nearest neighbors at the vertices. What drives lower symmetry is the problem of placing ten electrons along three perpendicular axes with six equivalent neighbors. Rhombohedral and ortho-rhombic distortions divide the six immediate neighbors in two distinct sets of first neighbors and second neighbors.

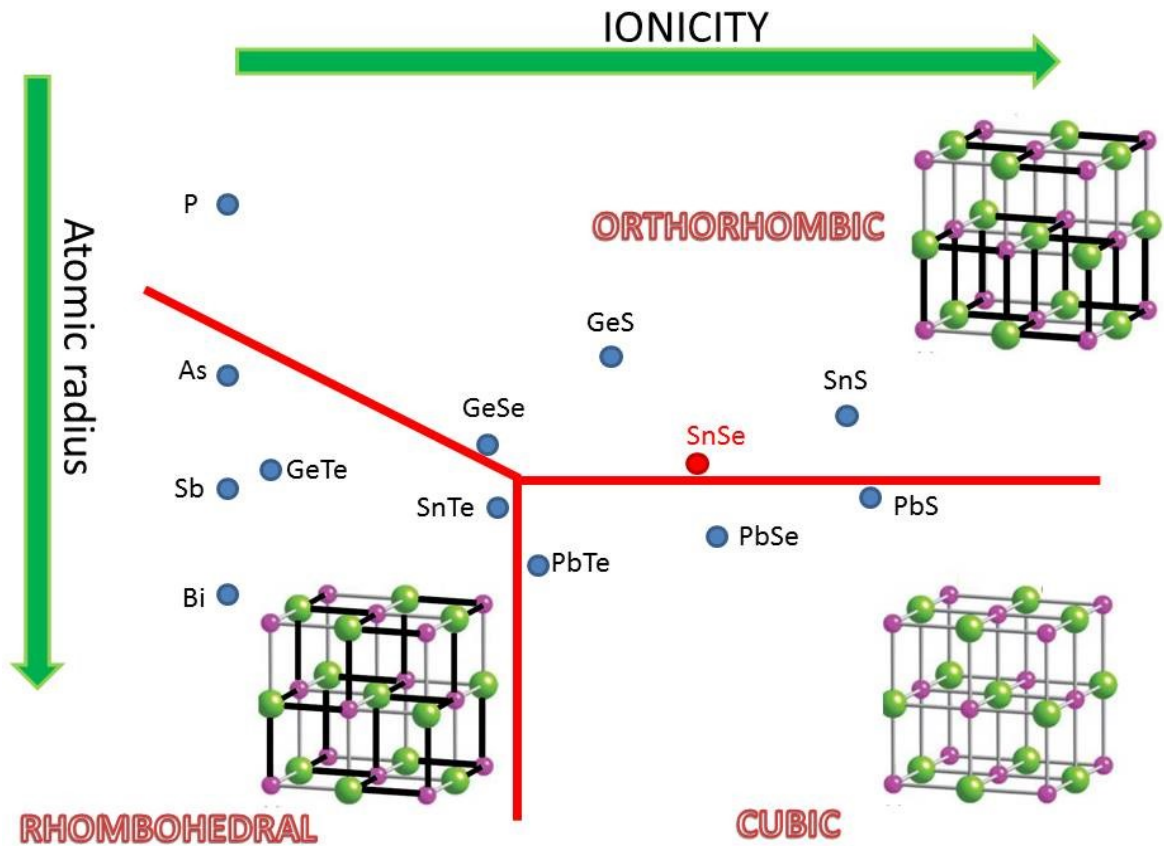

*Figure 1: IV-VI salts and column-V elements crystallize in three different structures, according to their size and iconicity. Both distortions of the symmetric cubic rocksalt structure favor three out of the six nearest neighbors. Solid black lines represent bonds strengthened by the shift of neighbors towards each other. In the orthorhombic structure of SnSe interlayer bonds are weak generating substantial anisotropy (Source: ref. 3 and 4).*

As seen in Fig.1, the heavier and ionic salts in the bottom–right corner of the diagram, like PbTe, avoid the distortion. This implies that the cubic symmetry is preserved in presence of strong spin-orbit interaction and large iconicity. Thanks to spin-orbit interaction, the six p and the four s orbitals become energetically distinct. With only p electrons at stake, it is much easier to keep six neighbors along three perpendicular axes. Moreover, ionicity, by distinguishing between partner atoms, impedes Pierels dimerization in each of the three perpendicular chains [5]. In short, covalency favors rhombohedral distortion, while s-p hybridization leads to orthorhombicity and a layered anisotropic structure (See Figure).

Is there any link between this structural competition and the remarkable thermoelectric figure of merit in SnSe and PbTe? One can identify at least two. The first concerns the unusually low lattice thermal conductivity. Its magnitude at 800 K

(in both PbTe and SnSe) implies that the phonon mean-free-path becomes as short as the interatomic distance, the lowest one may expect in a solid [6]. This is a result of strong phonon-phonon scattering, itself caused by the anharmonicity due to the proximity of the structural instability [7].

The second link is specific to SnSe in which ZT is large along one crystalline axis. This is because charge flows easily along the b-axis, yet the Seebeck coefficient is as large as along the other axes. The anisotropy of conductivity (as in the case of black phosphorus) is a consequence of the layered structure arising from orthorhombicity. Is it surprising that the Seebeck coefficient of this anisotropic conductor is almost isotropic? Not much, considering that in a simple anisotropic Fermi liquid, the Fermi radius and the thermal fuzziness of the Fermi surface share the same anisotropy and the Seebeck coefficient is set by the ratio of these two [6]. Indeed, the Seebeck coefficient remains isotropic in numerous layered conductors.

The present work will certainly inspire many other studies. Are other anisotropic conductors of the family as interesting? How different are the Fermi surface topologies among the family members? According to band calculations, a carrier concentration of 4 1019 cm-3, the Fermi surface of p-doped SnSe consists of several anisotropic pockets at low-symmetry-points [1]. This is to be checked by experimental study of quantum oscillations, as in the better-documented case of PbTe [8].

1. L.-D. Zhao et al., Science 350, xxxx (2015)
2. L.-D. Zhao et al., Nature 508, 373 (2014)
3. P. Littlewood, J. Phys. C: Solid State Phys. 13, 4855 (1980)
4. J. K. Burdett and T. J. McLarnan, Journal of Chemical Physics 75, 5764 (1981)
5. R. Peierls, More surprises in theoretical physics, Princeton University Press (1991)
6. K. Behnia, Fundamentals of thermoelectricity, Oxford (2015)
7. J. An, A. Subedi and D. J. Singh, Solid State Commun. 148, 417 (2008)
8. J. D. Jensen, B. Houston, and J. R. Burke, Phys. Rev. B 18, 5567 (1978)